\documentclass[11pt,a4paper]{article}
\usepackage[T1]{fontenc}
\usepackage[utf8]{inputenc}
\usepackage[english]{babel}
\usepackage[a4paper,margin=2.5cm]{geometry}
\usepackage{lmodern}
\usepackage{microtype}
\usepackage{amsmath,amssymb}
\usepackage{graphicx}
\usepackage{booktabs,tabularx,array,multirow}
\usepackage{pdflscape}
\usepackage{caption}
\usepackage{float}
\usepackage{setspace}
\usepackage[numbers,sort&compress]{natbib}
\usepackage{hyperref}
\hypersetup{colorlinks=true,linkcolor=blue,citecolor=blue,urlcolor=blue}
\newcommand{\PreserveBackslash}[1]{\let\temp=\\#1\let\\=\temp}
\newcolumntype{C}[1]{>{\PreserveBackslash\centering}m{#1}}
\newcolumntype{R}[1]{>{\PreserveBackslash\raggedleft}m{#1}}
\newcolumntype{L}[1]{>{\PreserveBackslash\raggedright}m{#1}}

\title{\bfseries KM3-230213A: The Highest-Energy Cosmic Neutrino}
\author{Carla Distefano\\[4pt]
\small Istituto Nazionale di Fisica Nucleare (INFN),\\
\small Laboratori Nazionali del Sud (LNS), Catania, Italy}
\date{}

\begin{document}
\maketitle
\begin{abstract} High-energy neutrinos provide a unique probe of the most extreme non-thermal processes in the Universe. The detection of the ultra-high-energy neutrino event KM3-230213A by the KM3NeT neutrino telescope, with an inferred parent-neutrino energy of $\mathcal{O}(10^{17}\,\mathrm{eV})$, represents a major step forward in neutrino astronomy. In this review, we summarize the main properties of the event and discuss its interpretation within different astrophysical and cosmological scenarios, including active galactic nuclei, transient sources, and cosmogenic neutrinos, highlighting the role of multi-messenger constraints. Despite extensive follow-up observations, no unambiguous electromagnetic counterpart has been identified, and the event remains in tension with the absence of comparable detections in other experiments. This situation leaves open the possibility of more complex source scenarios or contributions from physics beyond the Standard Model. We also discuss the implications for the ultra-high-energy neutrino landscape and the prospects for future observations with next-generation detectors. KM3-230213A provides a unique opportunity to probe particle acceleration and neutrino production at energies far beyond the reach of terrestrial experiments. \end{abstract}

\section{Introduction}

The detection of high-energy cosmic neutrinos has opened a new window in multi-messenger astronomy, providing direct access to the most extreme environments in the Universe \cite{halzen2017}. Unlike charged cosmic rays, neutrinos propagate undeflected by magnetic fields and can escape dense astrophysical regions, carrying unique information about the mechanisms of particle acceleration and hadronic interactions at the highest energies. The field of neutrino astronomy was established by the discovery of a diffuse flux of astrophysical neutrinos by the IceCube detector at the South Pole in 2013 \cite{evidence-he-nu,PhysRevLett.111.021103}, marking the first unambiguous observation of high-energy neutrinos of cosmic origin. Subsequent observations have provided the first compelling evidence for astrophysical neutrino sources, most notably the association with the blazar TXS~0506+056 \cite{ic:TXS0506+056}, as well as the identification of promising steady emitters such as the Seyfert galaxy NGC~1068 \cite{ic:NGC1068}. These results highlight the need for complementary detectors with different geographical locations and detection media, capable of improving sky coverage and enhancing the sensitivity to astrophysical neutrino sources.

The KM3NeT research infrastructure, currently under construction in the Mediterranean Sea, comprises two deep-sea neutrino detectors with distinct scientific goals: ORCA (Oscillation Research with Cosmics in the Abyss), located offshore Toulon, France, and primarily dedicated to neutrino oscillation studies, and ARCA (Astroparticle Research with Cosmics in the Abyss), located offshore Capo Passero, Sicily, and optimized for high-energy neutrino astronomy \cite{KM3NeTDetector}.
Located at a depth of about 3500~m, KM3NeT/ARCA is designed to explore the high-energy neutrino sky above the TeV scale. Thanks to its location in the Northern Hemisphere, KM3NeT provides a field of view complementary to that of IceCube, with enhanced sensitivity to the Galactic plane and the Galactic Centre. Although still under construction, KM3NeT/ARCA is already delivering significant physics results, demonstrating its capability for high-energy neutrino observations.

In February 2023, the KM3NeT/ARCA detector recorded an exceptional event, KM3-230213A, with a median parent-neutrino energy estimated at about 220~PeV \cite{Nature2025}. This detection represents the highest-energy neutrino ever observed and poses significant challenges for our understanding of ultra-high-energy (UHE) neutrino production. The extreme energy of the event, combined with the absence of comparable detections in other experiments, makes its origin particularly difficult to interpret within standard astrophysical scenarios.

In this work, we review the observational properties of KM3-230213A and discuss its implications in the broader context of multi-messenger astrophysics and fundamental physics. After presenting the main features of the event and the neutrino flux which can be inferred from it, we examine the constraints arising from comparisons with other neutrino experiments and explore the main classes of astrophysical and beyond-Standard-Model (BSM) interpretations proposed in the literature. The literature considered in this review includes works available up to April 2026.

\section{Observation of the KM3-230213A event}

\begin{figure}[h!]
\centering
\includegraphics[width=0.7\textwidth]{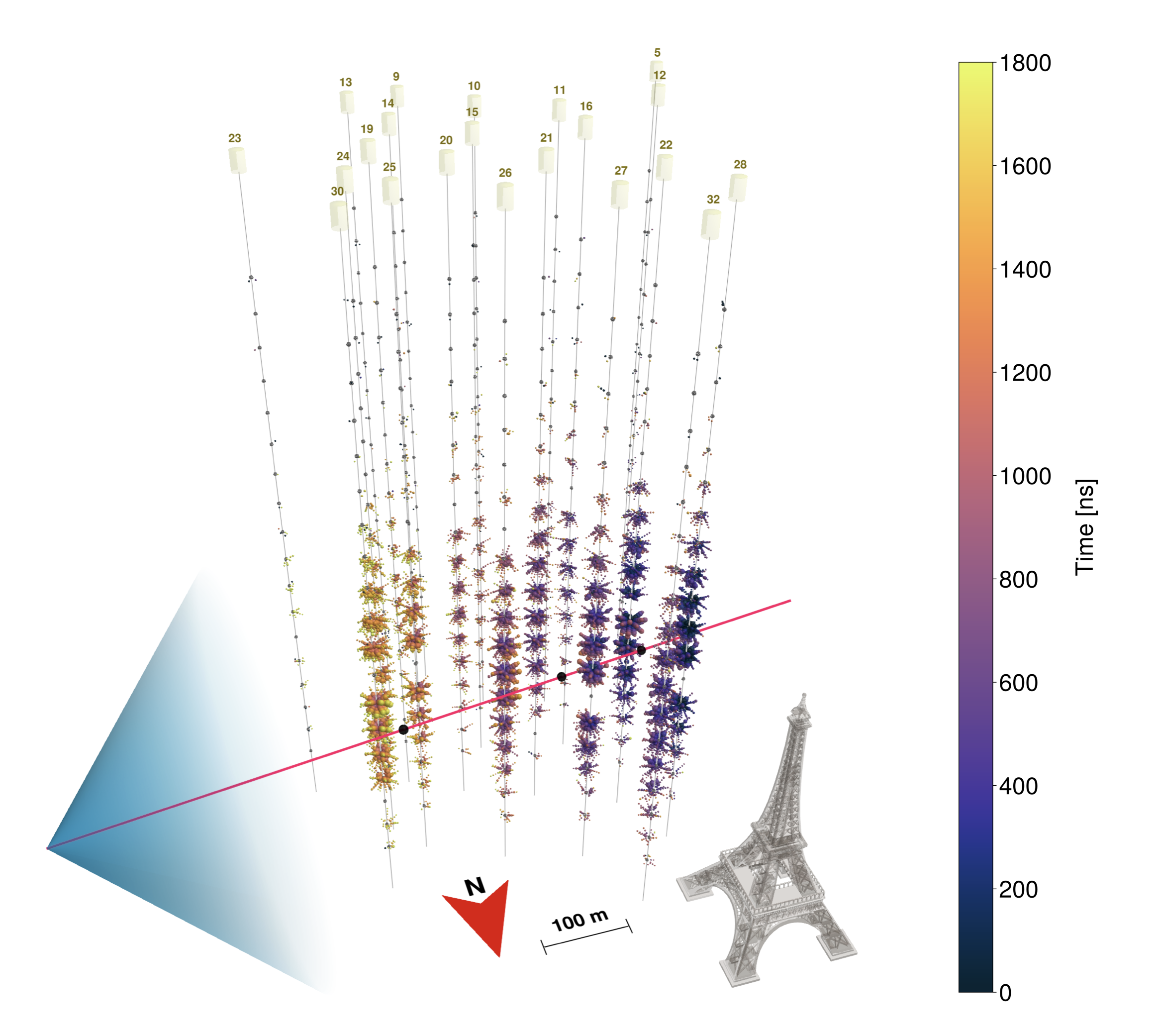}
\caption{Side view of the event KM3-230213A. The reconstructed muon track is shown together with an artist’s representation of the Cherenkov light cone. The PMT hits are colour-coded according to their detection time and scaled by the detected charge. Figure adapted from \cite{Nature2025}.}
\label{fig:eventdisplay}
\end{figure}

KM3-230213A was recorded on February 13, 2023, when the KM3NeT/ARCA detector comprised 21 operational detection lines. Over a total live time of 287.4~days in this detector configuration, about $1.1\times10^8$ events were triggered, among which one stood out due to its exceptional brightness and number of triggered photomultiplier tubes (PMTs). The event triggered $N_{\text{trig}} = 3672$ PMTs, clearly distinguishing it from the rest of the data sample \cite{Nature2025}. A visualization of the event is shown in Fig.~\ref{fig:eventdisplay}, illustrating the reconstructed muon track and the corresponding pattern of detected Cherenkov light.

\begin{figure}[h!]
\centering
\includegraphics[width=0.5\textwidth]{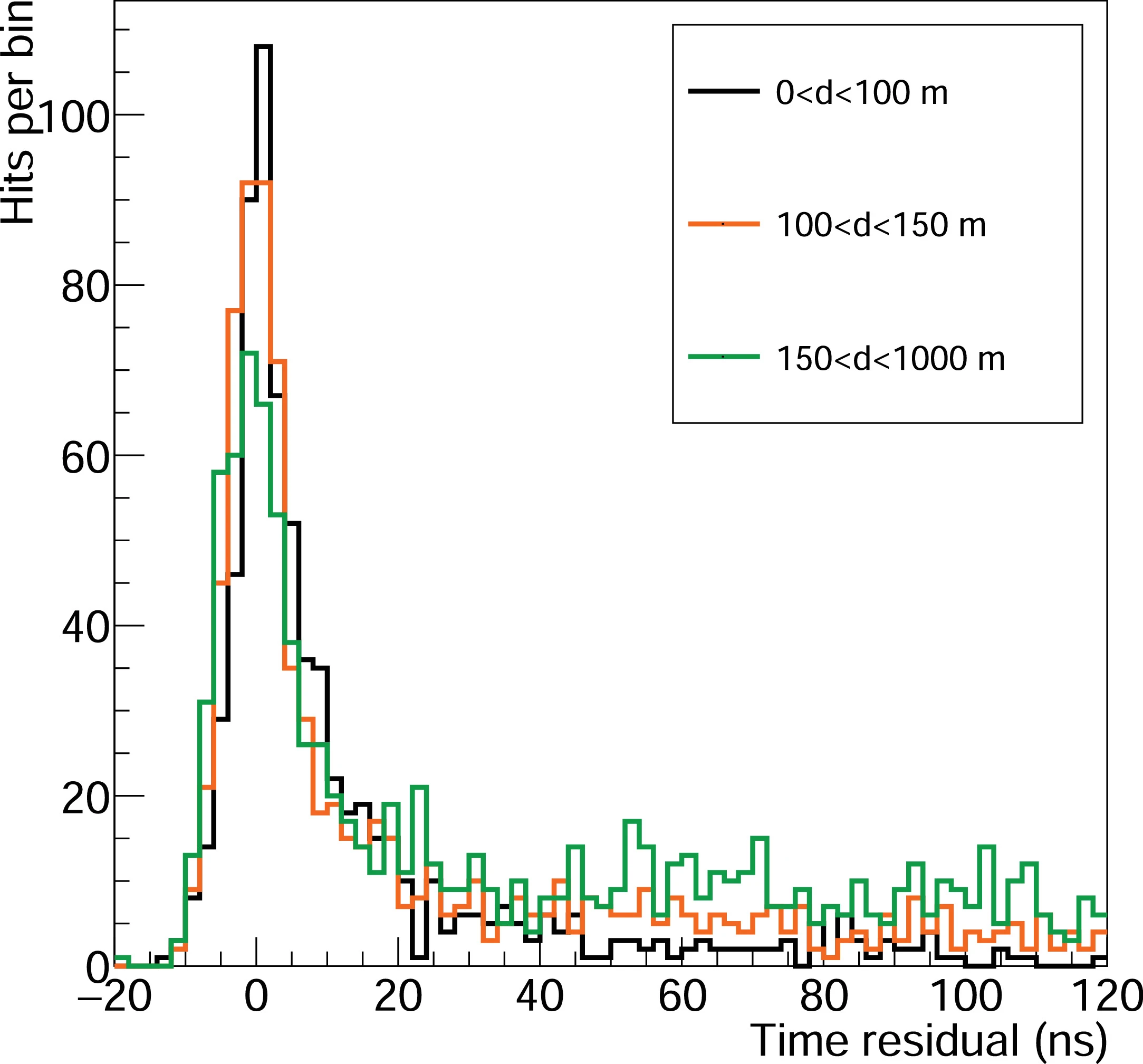}
\caption{Distribution of time residuals for KM3-230213A for different ranges of the distance $d$ from the reconstructed muon track. The characteristic peak around zero and positive tail are observed at different distances from the track, supporting the single-muon track hypothesis. Figure adapted from \cite{Nature2025}.}

\label{fig:timeresiduals}
\end{figure}

The spatial and temporal patterns of the recorded hits are fully compatible with the passage of a high-energy muon through the detector. The spatial distribution of the detected signals is consistent with a straight muon track, while their timing follows the expectation from Cherenkov emission. 
This is further confirmed by the distribution of the \textit{time residuals}, defined as the difference between measured and expected photon arrival times, where the expected arrival times are calculated assuming direct Cherenkov light from the muon. The distribution is sharply peaked around zero, as expected for direct Cherenkov photons, while the positive tail arises from photons that have scattered in the water and/or were emitted at angles different from the Cherenkov angle with respect to the muon track (Fig.~\ref{fig:timeresiduals}).

The event was reconstructed using a maximum-likelihood fit, yielding a near-horizontal track at $0.6^\circ$ above the horizon with an angular uncertainty of about $1.5^\circ$ (68\% C.L.), dominated by systematic uncertainties related to the absolute orientation of the detector. Future improvements in the detector calibration are expected to reduce these uncertainties toward the statistical limit.

\begin{figure}[h!]
\centering
\includegraphics[width=1\textwidth]{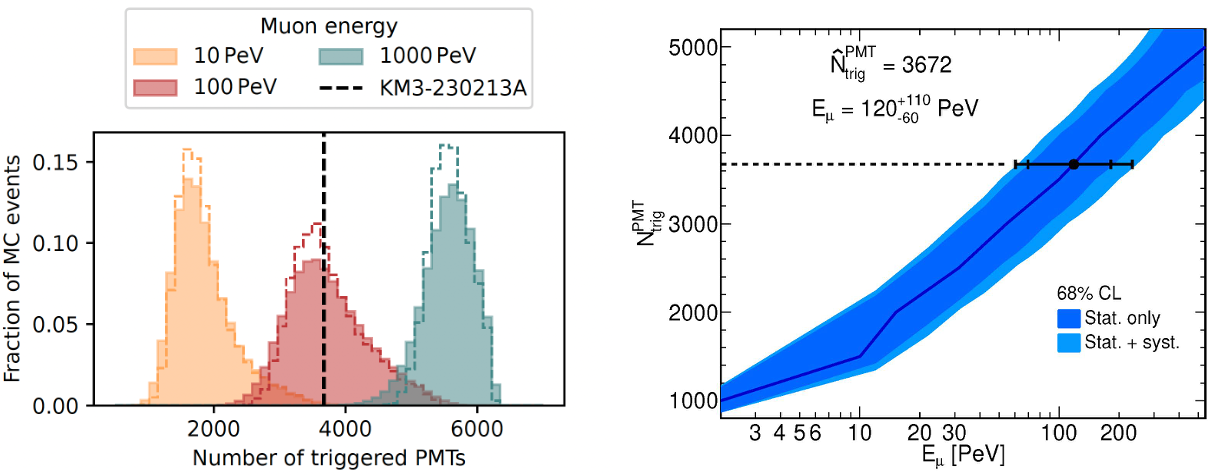}
\caption{Muon energy estimation for the KM3-230213A event: 
Left: Distribution of the number of triggered PMTs obtained from Monte Carlo simulations for different muon energies (10, 100, and 1000~PeV). The vertical dashed line indicates the observed value for KM3-230213A. 
Right: Reconstruction of the muon energy as a function of the number of triggered PMTs. The shaded bands represent the 68\% confidence level intervals including statistical uncertainties only (dark blue) and statistical plus systematic uncertainties (light blue). The horizontal dashed line shows the observed number of triggered PMTs, yielding an estimated muon energy of $E_\mu = 120^{+110}_{-60}$~PeV. Figures adapted from \cite{Nature2025}.}
\label{fig:muonenergy}
\end{figure}

Due to saturation effects in a significant fraction of PMTs close to the track, the standard energy reconstruction based on the measured light yield from muon energy losses could not be reliably used. Instead, a dedicated energy estimator based on the number of triggered PMTs was adopted to infer the muon energy at the detector from dedicated simulations. As illustrated in Fig.~\ref{fig:muonenergy}, the observed number of triggered PMTs is compared with simulated expectations for different muon energies, yielding an estimate of $E_\mu = 120^{+110}_{-60}$~PeV (68\% C.L.), with a 90\% confidence interval of 35--380~PeV.

The parent-neutrino energy cannot be inferred directly from the estimated muon energy, since the neutrino interaction vertex is not observed and neither the muon energy at production nor the energy lost by the muon before reaching the detector is known. It is therefore estimated using Monte Carlo simulations of neutrino interactions in a large volume surrounding the detector and the subsequent propagation of the produced muons. The simulated neutrino events are weighted assuming an $E_\nu^{-2}$ incident spectrum, and the parent-neutrino energy distribution is derived from events producing a muon at the detector compatible with the estimated muon energy of KM3-230213A. The assumed spectrum determines the relative weight assigned to the different parent-neutrino energies that can produce such muons and therefore affects the inferred parent-neutrino energy distribution, including its median and central range. For an $E_\nu^{-2}$ spectrum, this procedure yields a median parent-neutrino energy of about 220~PeV, with a central 90\% range of 72~PeV--2.6~EeV. This range should therefore not be interpreted as a spectrum-independent confidence interval on a directly reconstructed neutrino energy. For comparison, a previous well-known energy benchmark in neutrino astronomy was the IceCube Glashow-resonance event, with a reconstructed shower energy of $(6.05\pm0.72)$~PeV \cite{IceCubeGlashow}. Despite the model dependence associated with the inferred parent-neutrino energy, KM3-230213A represents the highest-energy neutrino event detected to date.

At such ultra-high energies, the Earth becomes largely opaque to neutrinos, strongly suppressing the flux from upgoing directions. The slightly downgoing, near-horizontal direction of KM3-230213A is therefore favorable for its detection, as it avoids propagation through large column depths of the Earth.

\subsection{Background estimation}

\begin{figure}[h!]
\centering
\includegraphics[width=1\textwidth]{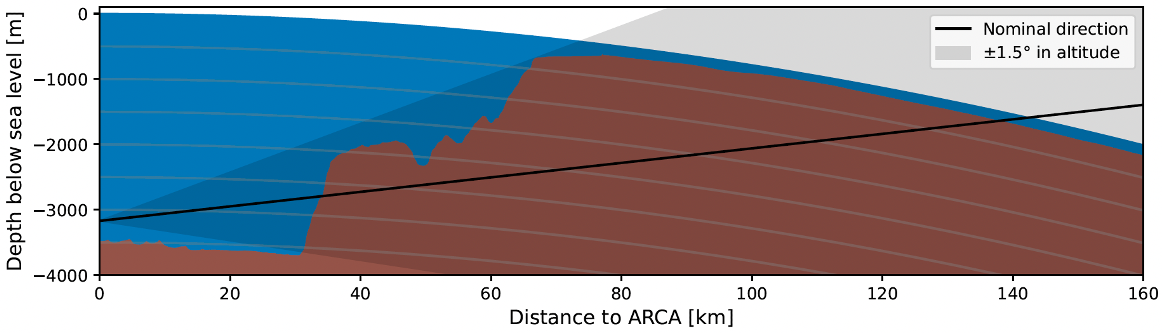}
\caption{
Topography along the reconstructed trajectory of KM3-230213A, showing the large amount of material traversed before reaching the detector. The shaded region shows the effect of a $\pm 1.5^\circ$ variation in the reconstructed zenith angle, corresponding to the 68\% systematic uncertainty. The corresponding path length, of several hundred kilometers water equivalent, strongly suppresses the atmospheric muon background. Figure adapted from \cite{Nature2025}.
}
\label{fig:malta_bathymetry}
\end{figure}

Given its reconstructed direction and energy, an atmospheric muon origin for KM3-230213A is strongly disfavored. The inferred trajectory implies that the particle traversed about 300~km water equivalent before reaching the detector. Even for a much higher muon energy of 100~EeV, the maximal range is only about 60~km water equivalent, still far below the column depth along the reconstructed direction. This is illustrated in Fig.~\ref{fig:malta_bathymetry}, which shows the topography along the event direction and the large amount of material traversed. Dedicated simulations yield an upper limit on the atmospheric-muon contamination of about $10^{-10}$ events per year at 100~PeV, even allowing for a shift of up to $2^\circ$ in the reconstructed zenith angle \cite{Nature2025}.

At such extreme energies, the conventional atmospheric-neutrino component, arising from pion and kaon decays, is strongly suppressed, fading away at energies of about 100~TeV--1~PeV. The atmospheric-neutrino flux at higher energies is therefore dominated by the prompt component produced in decays of short-lived charmed hadrons. However, the expected event rate above 100~PeV is of order $10^{-5}$ events per year, effectively excluding an atmospheric origin \cite{Nature2025}.

\subsection{Flux estimation and global neutrino landscape}

\begin{figure}[h!]
\centering
\includegraphics[width=0.85\textwidth]{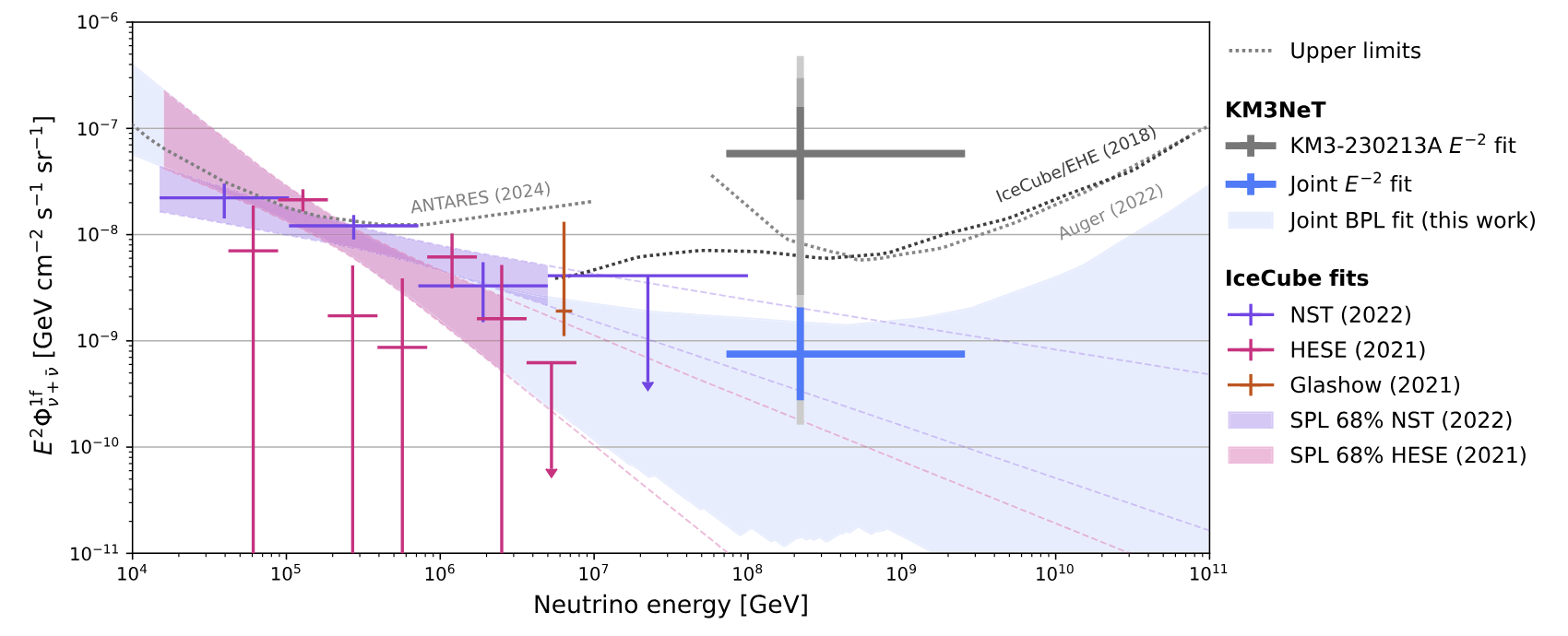}
\caption{Comparison of the diffuse neutrino flux inferred from the KM3-230213A event with measurements and upper limits from IceCube, Auger, and other experiments. The KM3NeT single-event estimate appears in tension with the absence of detections at comparable energies in other detectors, highlighting the current uncertainty in the UHE neutrino landscape. Figure adapted from \cite{km3net-landscape}, to which the label ``this work'' refers.}
\label{fig:neutrino_landscape}
\end{figure}

The standalone neutrino flux associated with KM3-230213A was derived using a total exposure of 335 days, combining data collected with 19 and 21 operational detection lines \cite{Nature2025}. This exposure differs from the 287.4-day live time quoted above, which refers specifically to the 21-line detector configuration used for the event sample. Using the central 90\% range of simulated parent-neutrino energies, 72~PeV--2.6~EeV, obtained under the $E_\nu^{-2}$ spectral assumption described above, a single observed event corresponds to

\begin{equation}
E_\nu^2 \, \Phi(E_\nu) = 5.8^{+10.1}_{-3.7} \times 10^{-8}~\mathrm{GeV~cm^{-2}~s^{-1}~sr^{-1}},
\end{equation}
as derived from KM3NeT data alone \cite{Nature2025}. This value is expressed as a per-flavor flux, assuming flavor equipartition at Earth, $\nu_e:\nu_\mu:\nu_\tau=1:1:1$. However, when combined with the absence of comparable events in IceCube and the Pierre Auger Observatory, a significantly lower flux normalization is preferred. A joint analysis within a global neutrino landscape framework yields

\begin{equation}
E_\nu^2 \, \Phi(E_\nu) = 7.5 \times 10^{-10}~\mathrm{GeV~cm^{-2}~s^{-1}~sr^{-1}},
\end{equation}
under the assumption of a steady isotropic flux described by a single power-law spectrum \cite{km3net-landscape}. This normalization is also given as a single-flavor flux. The comparison between the KM3NeT-only estimate and the global fit is illustrated in Fig.~\ref{fig:neutrino_landscape}. The discrepancy between these flux estimates reflects the tension between the KM3NeT observation and the absence of comparable events in other experiments. For the combined KM3NeT, IceCube UHE, and Auger datasets, assuming a steady isotropic flux described by a single power law, a parameter-goodness-of-fit test yields a $p$-value of 0.006, corresponding to a moderate tension of $2.5\sigma$ between the UHE datasets \cite{km3net-landscape}.

The same study also explored both single- and broken-power-law descriptions of the diffuse neutrino spectrum to test the possible emergence of an additional UHE component. While some IceCube datasets show a mild preference for a spectral break, no statistically significant evidence for such a component is found once all available constraints are included. Alternatively, the observed tension may reflect transient or anisotropic sources, for which the assumption of a steady isotropic flux does not apply.

Additional constraints arise from the connection between UHE neutrinos and their parent cosmic rays. The energy budget required to accommodate the KM3NeT event within standard photo-hadronic scenarios is extreme, while transient scenarios require both high luminosities and sufficiently abundant sources \cite{Das2025}. These considerations further emphasize the challenges in reconciling the KM3NeT observation with conventional astrophysical models.

\section{Interpretation and possible origins}

The origin of the UHE neutrino event KM3-230213A remains uncertain. With an inferred parent-neutrino energy of $\mathcal{O}(10^{17}\,\mathrm{eV})$, it is the highest-energy neutrino event detected to date. For comparison with the energy scales reached by other cosmic messengers, ultra-high-energy cosmic rays have been observed at energies exceeding $10^{20}\,\mathrm{eV}$ \cite{TA2023}. The exceptional energy of KM3-230213A places strong constraints on possible production scenarios without uniquely identifying its source. Current interpretations include a possible Galactic origin, active galactic nuclei (AGN), transient astrophysical sources such as gamma-ray bursts (GRBs), cosmogenic neutrinos produced during UHE cosmic-ray propagation, and scenarios involving physics beyond the Standard Model.

These scenarios differ substantially in their physical assumptions, observational constraints, and current level of support. Before discussing them individually, we summarize the main observational constraints relevant to their interpretation.

\subsection{Observational constraints}

Before considering possible interpretations of KM3-230213A, it is useful to
summarize the main observational constraints that any viable scenario must
satisfy. The event localization covers a relatively broad region of the sky,
within which extensive multi-wavelength searches have identified several
candidate sources but no statistically significant counterpart
\cite{Nature2025}. In particular, no significant
gamma-ray counterpart has been detected, while radio observations reveal a
crowded field containing a large number of sources, including several
candidate AGNs \cite{VERITAS,Crnogorevic2025,ASKAPradio}. No transient
electromagnetic counterpart has been firmly associated with the event.

At the same time, the flux inferred from KM3-230213A is constrained by the
absence of comparable UHE-neutrino detections in other experiments. IceCube
places stringent limits on a steady diffuse flux at these energies, while
constraints from the Pierre Auger Observatory further restrict scenarios
predicting a substantial diffuse UHE-neutrino component. As discussed in
Section~2, these observations create tension with a simple steady and
isotropic interpretation of the KM3NeT event and favor scenarios involving
transient or anisotropic emission, spectral features, or a suppressed
lower-energy flux. These observational constraints provide the framework
against which the different interpretations discussed below are assessed.
More detailed results from the electromagnetic follow-up campaigns are
presented in Section~4.

\subsection{Constraints on a Galactic origin}

A Galactic origin for KM3-230213A appears strongly disfavored. The reconstructed arrival direction ($l = 216.1^\circ, b = -11.1^\circ$) is offset by about $11^\circ$ from the Galactic plane. Nevertheless, this direction was considered compatible with the approximate extension of the Galactic interstellar medium in the initial KM3NeT analysis, motivating dedicated searches for Galactic counterparts \cite{Nature2025}. A dedicated analysis of the Galactic scenario has subsequently been presented in Ref.~\cite{GalacticOrigin}. At the position of the event, the expected contribution from standard Galactic diffuse neutrino emission at ultra-high energies is negligible and cannot account for the observed signal. Furthermore, no suitable Galactic accelerator has been identified near the reconstructed direction. For a photopion-production scenario, the observed neutrino would require parent cosmic rays with energies of at least approximately $4$~EeV per nucleon \cite{GalacticOrigin}, posing a significant challenge for conventional Galactic accelerators.

\begin{figure}[h!]
\centering
\includegraphics[width=1\textwidth]{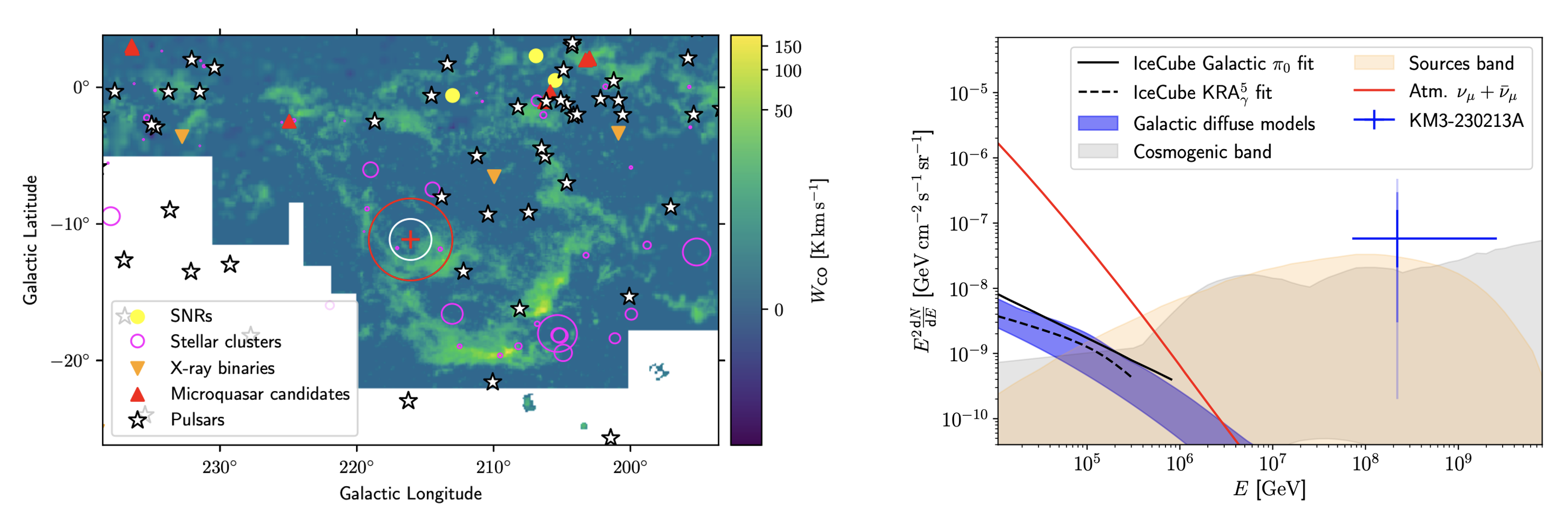}
\caption{Constraints on a Galactic origin of KM3-230213A. Left: reconstructed position of the event relative to the Galactic plane and potential Galactic accelerators, illustrating the absence of an obvious source association. Right: expected standard Galactic diffuse neutrino flux at ultra-high energies (blue band), compared with the expected flux from individual extragalactic sources (orange band) and cosmogenic neutrinos (grey band). The Galactic diffuse component cannot account for the observed event; this constraint does not by itself exclude an unusual Galactic point source or transient. Figure adapted from \cite{GalacticOrigin}.}
\label{fig:galactic-origin}
\end{figure}

These arguments are illustrated in Fig.~\ref{fig:galactic-origin}. The left panel shows the event direction relative to the Galactic plane and potential Galactic accelerators, while the right panel compares the event with the expected standard Galactic diffuse neutrino flux. The latter constrains the diffuse Galactic contribution but does not by itself exclude an unusual Galactic point source or transient. Taken together, the absence of a suitable nearby accelerator and the strong constraints on the standard diffuse Galactic component favor an extragalactic interpretation of the event.

\subsection{Active galactic nuclei and blazars}

\begin{figure}[htbp]
\centering
\includegraphics[width=0.75\textwidth]{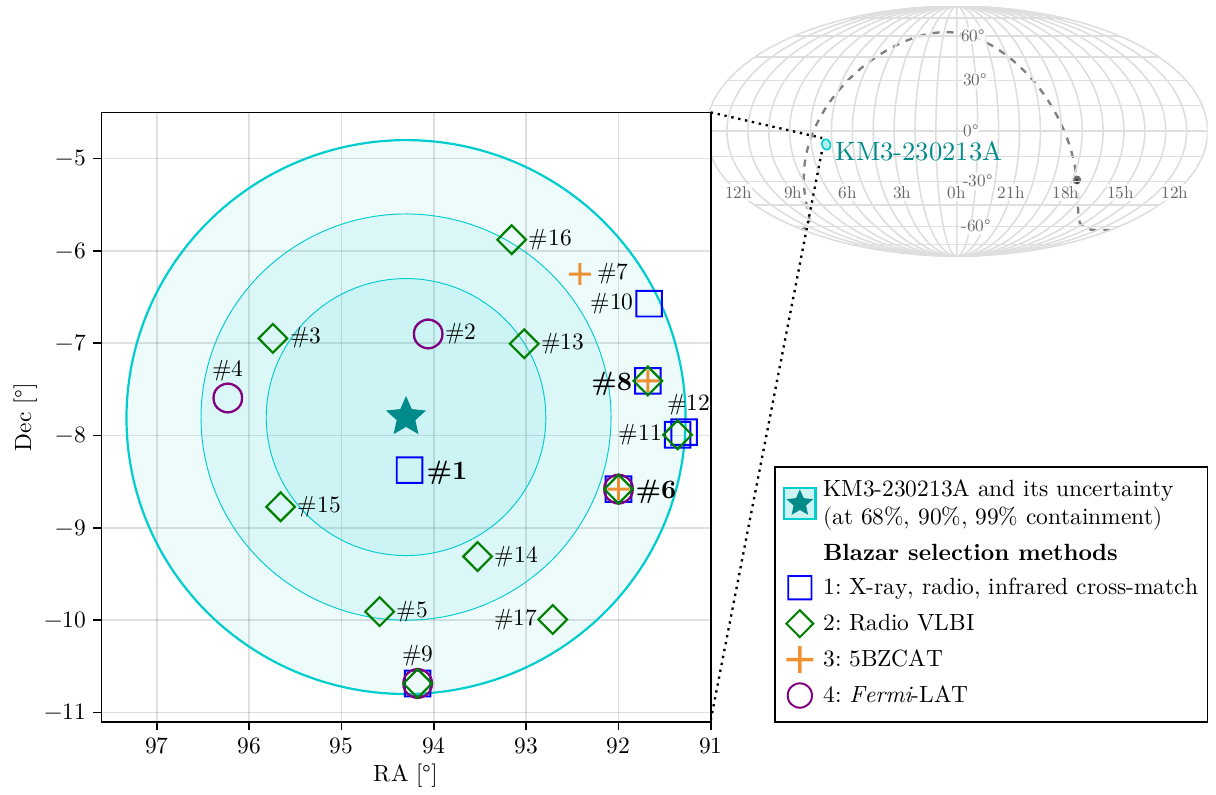}
\caption{Sky map of the KM3-230213A localization region and the 17 candidate blazar counterparts identified through multi-wavelength observations. The contours indicate the 68\%, 90\%, and 99\% containment regions of the reconstructed neutrino direction. The numbered sources correspond to the candidate blazars listed in Ref.~\cite{KM3NeTBlazarFollowup}, while the different markers indicate the selection method: X-ray/radio/infrared cross-matching, radio VLBI observations, the 5BZCAT catalogue, or Fermi-LAT observations. Sources \#1 (MRC 0614-083), \#6 (0605-085), and \#8 (PMN J0606$-$0724) are highlighted in the original study for detailed discussion. No statistically significant association with KM3-230213A has been established. Figure adapted from \cite{KM3NeTBlazarFollowup}.}
\label{fig:agn_skymap}
\end{figure}

Active galactic nuclei, and in particular blazars, are among the most widely discussed candidate sources of high-energy neutrinos. In these systems, particles can be accelerated to ultra-high energies within relativistic jets, with neutrino production occurring through photo-hadronic ($p\gamma$) or hadronuclear ($pp$) interactions \cite{halzen2017, murase2014blazars}. Producing neutrinos at $\sim 10^{17}$--$10^{18}$ eV, however, requires extreme acceleration conditions, close to the limits of standard jet models. Multi-wavelength searches within the localization region of KM3-230213A have identified 17 candidate blazars, with potentially interesting radio, X-ray, or gamma-ray activity reported for several of them \cite{Nature2025,KM3NeTBlazarFollowup}. However, none of these candidates or episodes of activity provides a statistically significant or physically compelling association with KM3-230213A. The candidate counterparts and event localization are shown in Fig.~\ref{fig:agn_skymap}.

\begin{figure}[htbp]
\centering
\includegraphics[width=0.85\textwidth]{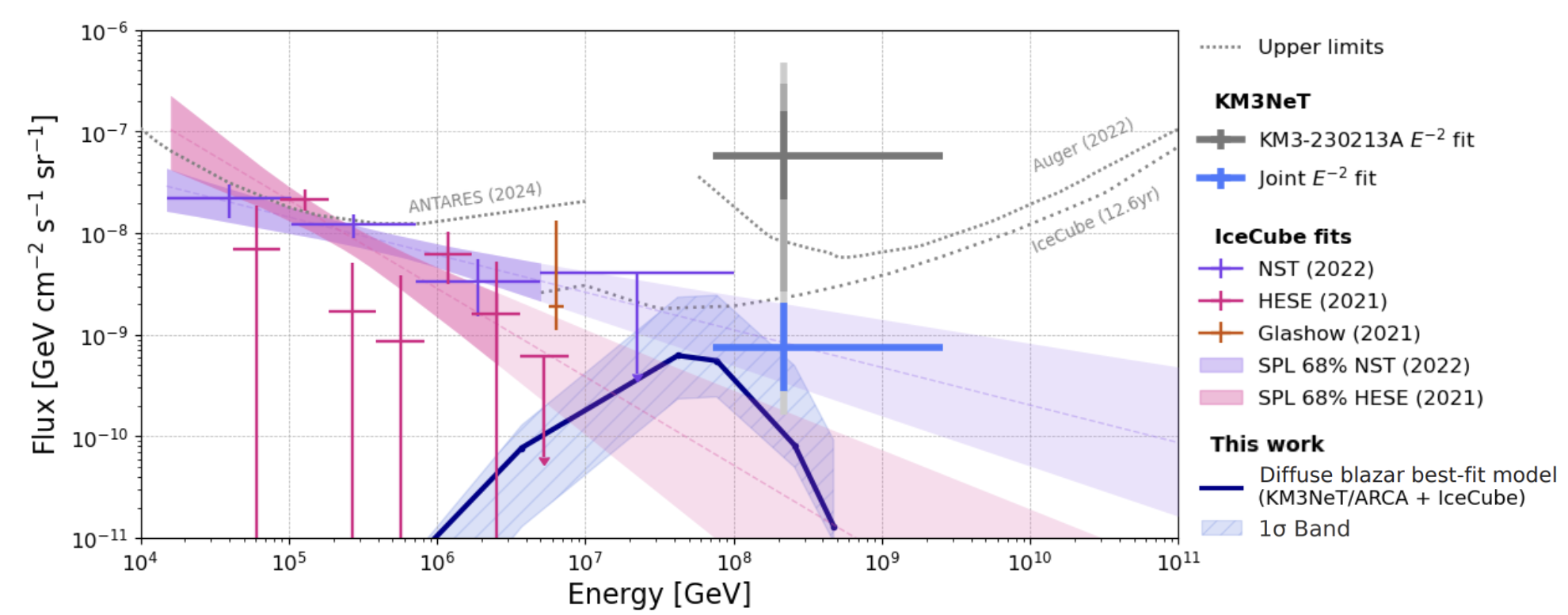}
\caption{Diffuse single-flavor neutrino flux predicted for the blazar population in the joint KM3NeT/ARCA--IceCube fit, compared with the KM3-230213A equivalent flux, IceCube measurements, and IceCube and Auger upper limits. Under this joint-fit assumption, the predicted flux is compatible with the experimental constraints and with the flux implied by KM3-230213A. Figure adapted from \cite{KM3NeTBlazars}.}
\label{fig:blazar_flux}
\end{figure}

A complementary approach considers the diffuse neutrino emission from the overall blazar population, rather than from any of the individual candidate counterparts discussed above \cite{KM3NeTBlazars}. Using lepto-hadronic models constrained by multi-messenger observations, the joint KM3NeT/ARCA--IceCube fit favors a baryonic loading $\eta \sim 10$ and a proton spectral index $\alpha_p \sim 1.8$. Under this joint-fit assumption, the predicted diffuse single-flavor neutrino flux is compatible with the flux implied by KM3-230213A and with existing IceCube and Auger constraints, as shown in Fig.~\ref{fig:blazar_flux}. In contrast, an analysis based on the KM3NeT/ARCA exposure alone requires a much larger baryonic loading, $\eta \sim 10^3$, with $\alpha_p \sim 1.75$, corresponding to an increase of about two orders of magnitude in the required proton luminosity. The resulting KM3NeT/ARCA-only diffuse neutrino spectrum is in tension with existing IceCube measurements and upper limits.

A number of works have explored more specific scenarios involving individual blazars within the localization region, including PKS 0605-085, MRC 0614-083, and PMN J0606$-$0724 \cite{PKS0605,MRC0614Paper,Kivokurtseva2025,JetRedGiant}. None of these candidates, however, can currently be regarded as a preferred counterpart. These interpretations typically invoke enhanced radiation fields, accretion-driven flares, or interactions between the relativistic jet and the surrounding environment to enable efficient $p\gamma$ interactions at ultra-high energies. In some cases, the associated gamma-ray emission is suppressed through internal absorption and electromagnetic cascades, allowing compatibility with multi-wavelength constraints. While such models can produce neutrinos with energies compatible with that inferred for KM3-230213A under favorable conditions, they generally require extreme source parameters and remain highly model dependent.

Overall, AGNs and blazars remain viable candidate sources of UHE neutrinos, and population models can accommodate KM3-230213A within constrained regions of parameter space. However, no individual blazar has been significantly associated with the event, and models invoking specific candidate counterparts generally require favorable or extreme source conditions. The absence of a statistically significant counterpart, together with stringent multi-messenger constraints, therefore prevents a definitive identification.

\subsection{Gamma-Ray Bursts and transient sources}

Gamma-ray bursts represent a well-established class of transient sources capable of accelerating particles to extreme energies. In the standard fireball framework, internal or external shocks in ultra-relativistic jets can produce high-energy neutrinos through photo-hadronic ($p\gamma$) interactions \cite{waxman1997high, meszaros2006grb_neutrinos}. The transient nature of GRBs provides a possible way to reconcile an isolated UHE event such as KM3-230213A with the absence of comparable detections over longer exposures. However, no GRB counterpart has been identified in temporal or directional coincidence with the event, and current gamma-ray observations provide no evidence for an electromagnetically bright transient within the reconstructed arrival direction.

\begin{figure}[htbp]
\centering
\includegraphics[width=0.85\textwidth]{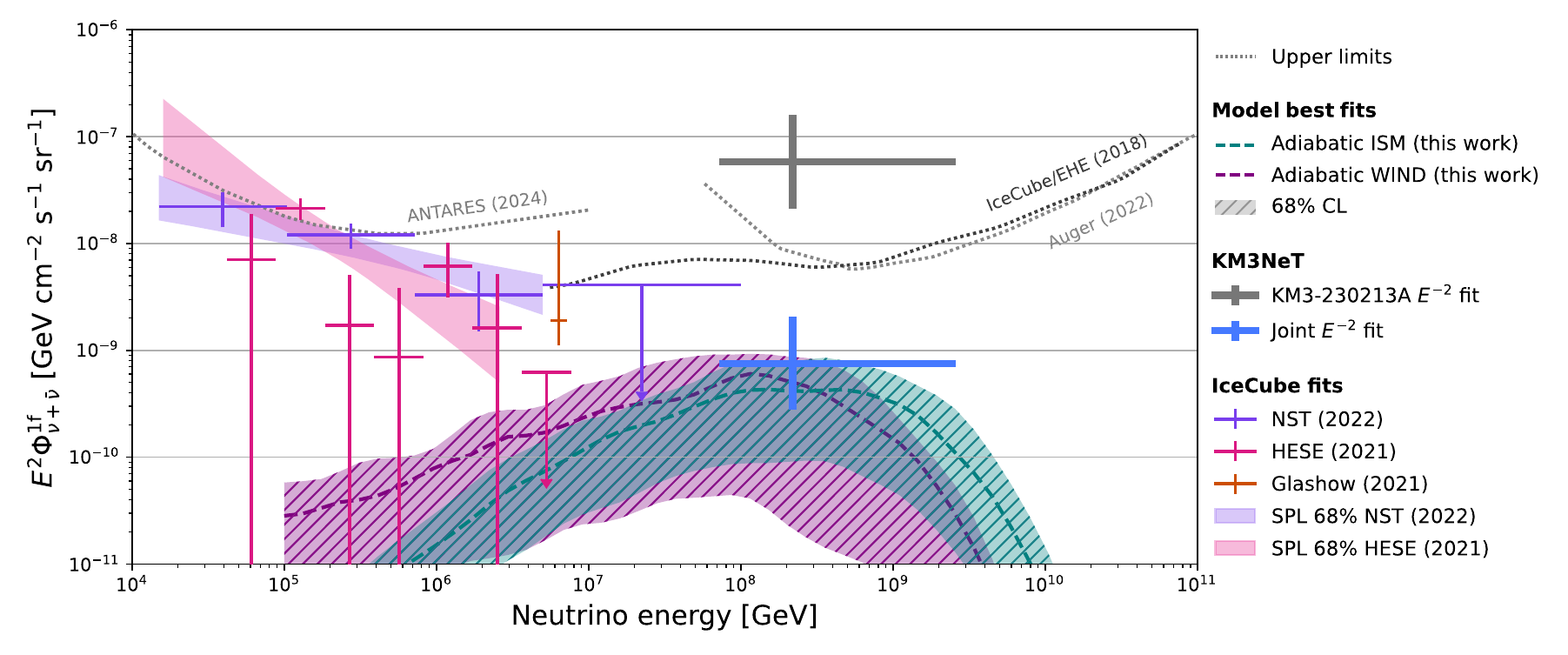}
\caption{Diffuse single-flavor neutrino flux from a population of long-GRB blastwaves, assuming flavor equipartition at Earth, compared with the KM3-230213A flux estimate and experimental measurements and constraints. The ISM and WIND curves correspond, respectively, to blastwaves expanding in a constant-density interstellar medium and in a wind-like environment with density decreasing as $R^{-2}$. The associated shaded regions show the 68\% confidence intervals of the predicted flux obtained from the model parameters constrained using the combined KM3NeT/ARCA, IceCube-EHE, and Auger exposure. Neutrinos are assumed to originate from photohadronic interactions of accelerated protons with afterglow photons. Figure adapted from \cite{KM3NeTGRB}.}
\label{fig:grb_flux}
\end{figure}

The KM3NeT Collaboration explored the compatibility of the event with neutrino production in GRB blastwaves, focusing on a cosmological population of long-duration bursts \cite{KM3NeTGRB}. In this scenario, UHE neutrinos are predominantly produced during the afterglow phase, as accelerated protons interact with synchrotron photons in the expanding blastwave. By combining the KM3NeT detection with the absence of comparable events in IceCube and the Pierre Auger Observatory, constraints are placed on key physical parameters such as the baryon loading and the density of the surrounding medium. These results constrain the allowed GRB parameter space when the KM3NeT detection is considered together with the non-observation of comparable events by IceCube and Auger. As shown in Fig.~\ref{fig:grb_flux}, the resulting GRB flux predictions approach the combined constraint but remain well below the KM3NeT single-event flux estimate.

\begin{figure}[htbp]
\centering
\includegraphics[width=0.65\textwidth]{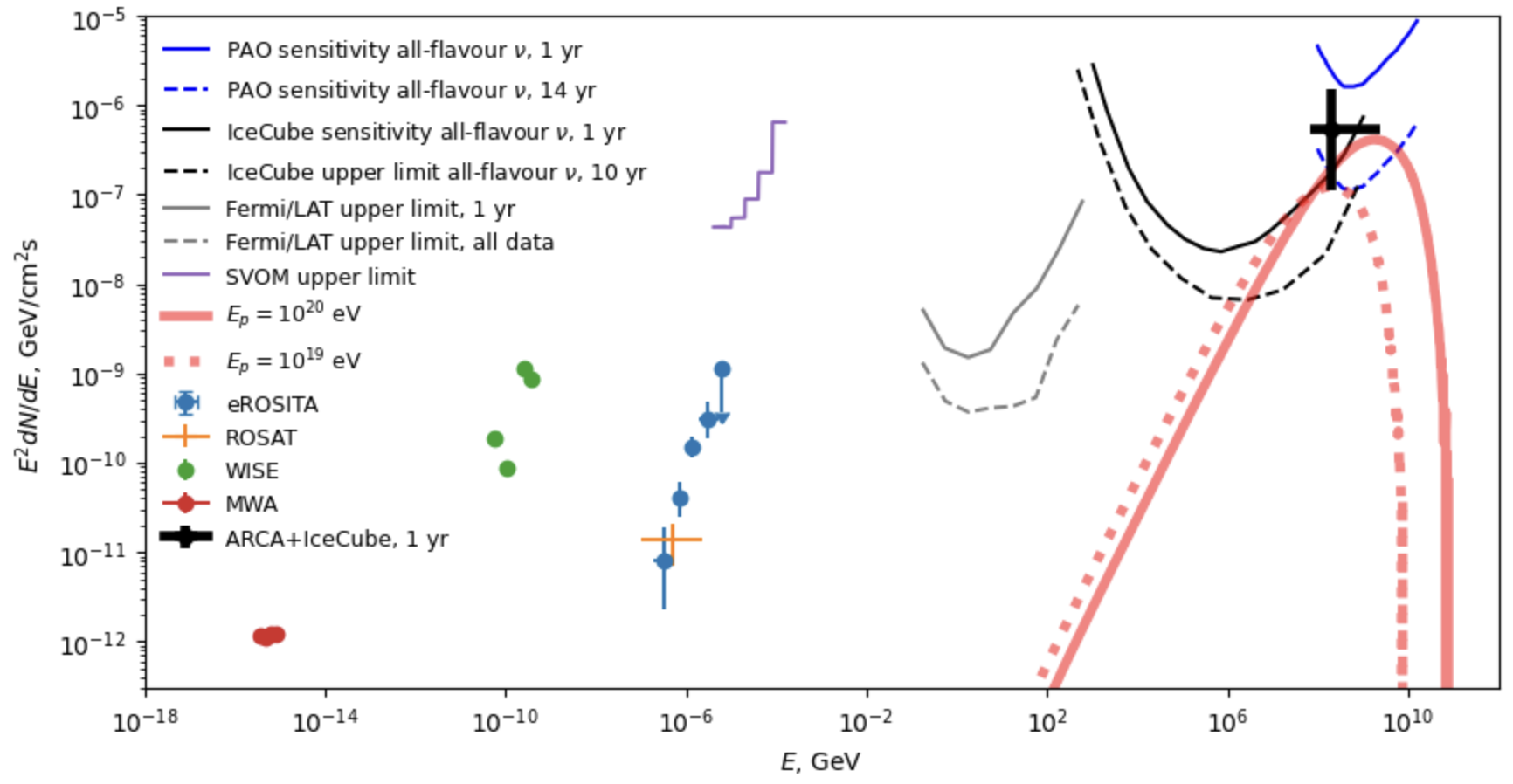}
\caption{Spectral energy distribution of TXS 0614-083, the AGN closest to the KM3-230213A direction, compared with the neutrino flux required in the isolated-source flare scenario. The black data point shows the estimated muon-neutrino flux for a one-year flare, obtained by combining the KM3NeT/ARCA detection with the IceCube non-detection; its uncertainty corresponds to the 90\% confidence interval. Radio, infrared, and X-ray measurements are from GLEAM, WISE, and eROSITA, respectively, while upper limits from SVOM and Fermi-LAT constrain the electromagnetic emission. IceCube sensitivities and the one-year and 14-year Pierre Auger Observatory sensitivities are also shown. The red curves illustrate secondary-neutrino spectra produced by interactions of $10^{19}$ and $10^{20}$ eV protons and demonstrate that a hard neutrino spectrum is required to avoid overproduction of lower-energy neutrinos. The transient interpretation assumes an isolated source flare with duration $T\lesssim2$ yr. Figure adapted from \cite{Neronov2025}.}

\label{fig:transient_sed}
\end{figure}

More generally, the event has been interpreted within a transient-source framework in which the neutrino is produced during a short-lived flare rather than by a steady emitter \cite{Neronov2025}. In this picture, the KM3NeT detection can be reconciled with the absence of similar events in other detectors if the emission occurs over a timescale $T \lesssim 2$ years and temporarily exceeds the diffuse background. The required energetics, of order $E \sim 10^{50}$ erg for cosmological distances, are compatible with powerful transients such as GRBs, tidal disruption events, or flaring active galactic nuclei (Fig.~\ref{fig:transient_sed}).

Several specific classes of transients have been proposed as potential sources. Low-luminosity GRBs, particularly those associated with shock breakout events, can produce multi-component neutrino spectra extending to ultra-high energies, with the afterglow component reaching $\sim 100$ PeV \cite{LL-GRBs}. Similarly, magnetar-powered optical transients, such as luminous fast blue optical transients (LFBOTs), have been suggested as efficient sites of cosmic-ray acceleration and neutrino production \cite{LFBOTs}. These models illustrate possible transient environments in which UHE-neutrino production may occur, although their applicability to KM3-230213A remains model dependent and no observational association with the event has been established.

GRBs and other transient sources therefore provide a physically motivated framework for UHE-neutrino production and can alleviate the tension between KM3-230213A and the absence of comparable events in other detectors. However, the lack of a clear temporal and spatial counterpart and the existing multi-messenger constraints prevent the identification of any specific transient class as a favored explanation.

\subsection{Cosmogenic neutrino interpretation}

One possible interpretation of KM3-230213A is that it belongs to the cosmogenic neutrino flux, produced by interactions of ultra-high-energy cosmic rays (UHECRs) with background photon fields during their propagation across cosmological distances. In this scenario, protons with energies above $\sim 10^{19}\,\mathrm{eV}$ interact with the cosmic microwave background (CMB) via photo-pion production, leading to charged pion decay and neutrino emission. This mechanism, known as the Greisen--Zatsepin--Kuzmin (GZK) process, naturally produces a diffuse neutrino flux extending up to $\sim 10^{18}$--$10^{19}\,\mathrm{eV}$ \cite{berezinsky1969cmb, anchordoqui2018cosmogenic}. Predictions of the cosmogenic neutrino flux remain subject to significant uncertainties, reflecting our limited knowledge of the properties of UHECR sources. Key parameters include the source emissivity, maximum acceleration energy, injection spectral index, cosmological evolution, and mass composition, all of which strongly affect the expected flux and spectral shape. The detection of cosmogenic neutrinos is therefore particularly interesting, as it would provide direct insight into the origin, composition, and propagation of UHECRs.

\begin{figure}[htbp]
\centering
\includegraphics[width=1\textwidth]{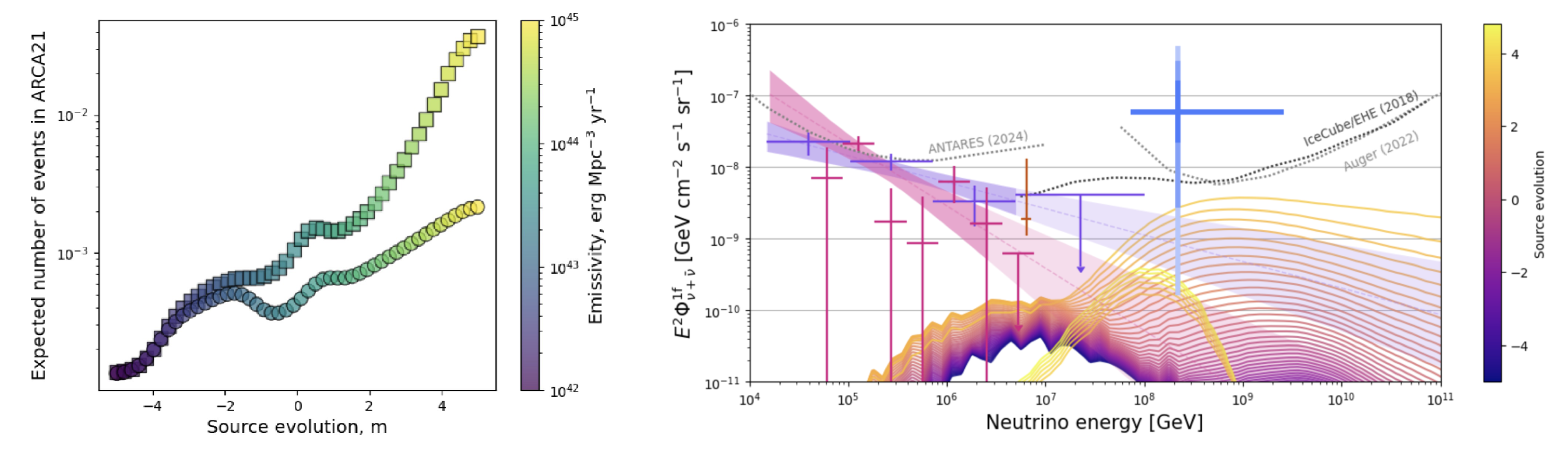}
\caption{Cosmogenic-neutrino predictions as a function of UHECR source evolution and composition. Left: expected number of events in ARCA21 as a function of the source-evolution parameter $m$, defined through $S(z)\propto(1+z)^m$, for source populations extending up to $z_{\max}=1$ (circles) and $z_{\max}=6$ (squares); the colour scale indicates the UHECR source emissivity. The expected event rate increases for stronger positive source evolution and for populations extending to higher redshift. Right: diffuse single-flavor cosmogenic neutrino fluxes for different values of $m$, including an additional subdominant proton component at the highest UHECR energies; stronger positive source evolution enhances the predicted flux. The blue cross represents the KM3-230213A equivalent flux, with the horizontal bar indicating the central 90\% parent-neutrino energy range and the vertical bars the 1, 2, and 3$\sigma$ Feldman--Cousins intervals. The purple and pink bands show the 68\% confidence intervals of IceCube single-power-law fits; measurements and upper limits from ANTARES, IceCube, and the Pierre Auger Observatory are also shown. Compatibility with KM3-230213A is enhanced for a source population extending to high redshift, strong positive source evolution, and a non-negligible proton fraction at the highest energies. Figure adapted from \cite{CosmogenicKM3NeT}.}

\label{fig:cosmogenic_figures}
\end{figure}

With an inferred parent-neutrino energy of $\mathcal{O}(10^{17}\,\mathrm{eV})$ under the $E_\nu^{-2}$ spectral assumption discussed in Section~2, KM3-230213A falls within the expected cosmogenic energy range, although standard models typically predict low event rates at these energies for current detector exposures. This motivated a dedicated study by the KM3NeT Collaboration assessing the compatibility of the event with cosmogenic scenarios under different assumptions on composition and source evolution \cite{CosmogenicKM3NeT}. Compatibility improves in models with strong positive evolution ($S(z)\propto(1+z)^m$ with $m>3$), contributions up to redshift $z_{\rm max}\simeq6$, and a non-negligible proton fraction at the highest energies (Fig.~\ref{fig:cosmogenic_figures}).

Beyond this analysis, several recent works have explored updated cosmogenic predictions incorporating improved treatments of UHECR composition and propagation. Models based on Telescope Array data with a predominantly light composition and standard cosmological evolution \cite{CosmogenicMinimalModels} predict a neutrino flux below, but compatible with, the KM3NeT measurement at the $\sim 2\sigma$ level, while also implying the expectation of $\mathcal{O}(1)$ events in other detectors, which are not observed. A complementary approach considers a two-population UHECR framework consisting of a dominant mixed-composition component and a subdominant proton population \cite{CosmogenicImplications}. In this case, the proton fraction is constrained to $\sim 20\%$ at $\sim 20$ EeV, whereas the required cosmological evolution depends strongly on the interpretation of the KM3NeT event: considering only KM3NeT favors strong evolution, while including IceCube and Auger limits allows for flatter or even negative evolution.

The role of extragalactic magnetic fields has also been investigated in scenarios involving transient UHECR sources \cite{CosmogenicPointSource,CosmogenicDelayedGamma}. Magnetic deflections affect both the temporal and angular distributions of cosmogenic neutrinos, potentially leading to point-like signals with time-dependent fluxes. Weak magnetic fields produce brighter but shorter-lived signals, whereas stronger fields result in a dimmer and more extended emission. In addition, the associated $\gamma$-ray emission is expected to be significantly delayed ($\sim 10^{4}$--$10^{6}$ years) due to cascade development and magnetic deflections, naturally explaining the absence of electromagnetic counterparts.

A global multi-messenger analysis further constrains cosmogenic scenarios by jointly modeling cosmic rays, neutrinos, and gamma rays \cite{CosmogenicMultimess}. These results disfavor steep proton spectra, which tend to overproduce the diffuse gamma-ray background, and instead support hard spectra extending beyond $10^{20}$ eV with moderate cosmological evolution and a relatively low proton fraction ($\lesssim 10\%$).

Overall, the cosmogenic interpretation provides a natural diffuse extragalactic framework for KM3-230213A, consistent with the absence of a clear point-source counterpart. Its compatibility with the event is, however, strongly model dependent and sensitive to assumptions on the UHECR composition, proton fraction, injection spectrum, source evolution, and, for transient scenarios, extragalactic magnetic fields. Joint constraints from IceCube, Auger, and gamma-ray observations substantially restrict the viable parameter space.

\subsection{Exotic physics and BSM scenarios}

The extreme energy of KM3-230213A opens the possibility of testing scenarios involving physics beyond the Standard Model. These include modifications of neutrino propagation, non-standard interactions, dark matter decay, primordial black holes, and more speculative scenarios involving new particles or unconventional production mechanisms \cite{bhattacharjee2000uhecr_review, feng2010uhe_neutrinos}. These scenarios play qualitatively different roles in the interpretation of KM3-230213A. In some cases, most notably Lorentz-invariance violation, the event primarily places constraints on new physics rather than providing evidence for it. Other models can accommodate the event only within restricted or highly model-dependent regions of parameter space and may predict additional testable signatures. The main assumptions, observational constraints, testable signatures, and role of KM3-230213A for the different BSM scenarios are compared systematically in Table~\ref{tab:scenario_comparison}. At present, none provides positive evidence for a BSM origin of KM3-230213A.

\subsubsection{Modified neutrino propagation and Lorentz invariance violation}

\begin{figure}[htbp]
\centering
\includegraphics[width=0.6\textwidth]{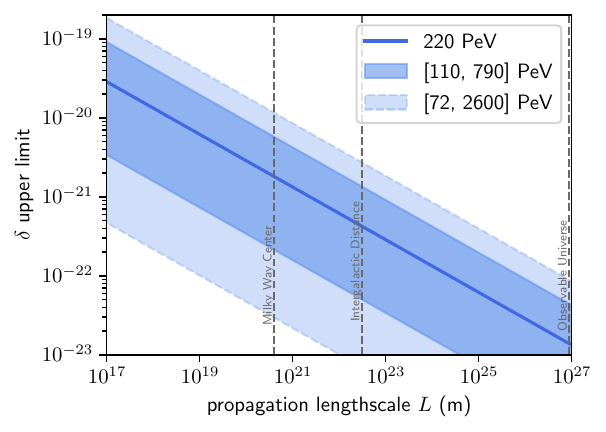}
\caption{Upper limits on the Lorentz-invariance-violation parameter $\delta\equiv c_\nu^2-1$ as a function of the neutrino propagation distance $L$, derived from KM3-230213A. The central curve assumes a parent-neutrino energy of 220~PeV, while the shaded bands reflect the central 68\% and 90\% ranges of the parent-neutrino energy distribution. Vertical dashed lines indicate representative propagation scales corresponding to the size of the Milky Way, an intergalactic distance of 1~Mpc, and the size of the observable Universe. The limits assume superluminal-neutrino energy losses through electron--positron pair emission, with non-superluminal electrons, and adopt a propagation distance equal to ten neutrino decay lengths. Figure adapted from \cite{km3net-lvi}.}
\label{fig:liv_constraint}
\end{figure}

KM3-230213A also provides a powerful probe of possible modifications of neutrino propagation over cosmological distances. In particular, the KM3NeT Collaboration has used the detection of KM3-230213A to set the most stringent constraints to date on Lorentz invariance violation (LIV) in the neutrino sector \cite{km3net-lvi}. In superluminal scenarios, UHE neutrinos would undergo rapid energy losses through processes such as electron--positron pair emission, preventing them from reaching Earth with energies as high as observed. The observation of a neutrino with an energy above 100 PeV therefore implies a stringent upper bound on the LIV parameter, significantly improving previous limits (Fig.~\ref{fig:liv_constraint}). Additional studies have explored complementary approaches, including neutrino decay, modified dispersion relations, Lorentz-violating effects in the muon sector, and possible associations with GRBs \cite{Satunin2025,Yang2025,Carmona2026,Amelino2025,Cattaneo2025,Wang2025,Wang2026}. These interpretations remain statistically inconclusive due to uncertainties in source associations. Thus, rather than providing evidence for LIV, KM3-230213A primarily places stringent constraints on superluminal Lorentz-violating propagation.

An alternative class of scenarios considers new physics affecting neutrino propagation. In models involving sterile neutrinos, the flux arriving at Earth may include a significant sterile component that partially converts into active neutrinos through matter-induced oscillations. Since this conversion depends on the propagation path through the Earth, different detectors may observe different effective fluxes, providing a characteristic path-dependent signature that could potentially alleviate the tension between KM3NeT and other experiments \cite{Brdar2026}. Such an interpretation requires a significant sterile-neutrino component and non-standard matter effects and therefore remains highly speculative.

\subsubsection{Neutrino non-standard interactions}

Beyond LIV, a broader class of models involves non-standard neutrino interactions that can affect the observed neutrino flux. Neutrino self-interactions mediated by light particles can produce characteristic spectral distortions and partially alleviate the tension between KM3NeT and IceCube \cite{He2026,Petropavlova2025}. However, UHE-neutrino observations place strong constraints on the relevant parameter space, limiting the extent to which such interactions can reconcile the two measurements. Interactions between neutrinos and dark matter along the line of sight provide another possibility, inducing energy-dependent attenuation effects that can modify the observed spectrum \cite{Bertolez2026,Mondol2025}. These scenarios require new particles and interactions beyond the Standard Model and remain highly model dependent. KM3-230213A can therefore constrain such interactions and, within restricted parameter regions, they may modify the neutrino spectrum, but the event provides no positive evidence for this scenario.

\subsubsection{Dark matter scenarios}

Another class of proposed interpretations involves the decay of super-heavy dark matter particles. In these models, dark matter with masses well above the PeV scale can decay into Standard Model particles, producing UHE neutrinos without requiring astrophysical acceleration mechanisms \cite{Borah2025,Kohri2025}. More elaborate realizations embed this idea within extended particle physics frameworks, including right-handed neutrino sectors, $U(1)$ extensions, and freeze-in production mechanisms \cite{Barman2025,Khan2025,Murase205,Su2025}. Multi-messenger constraints from neutrino and gamma-ray observations typically require extremely long dark matter lifetimes ($\gtrsim 10^{29}$--$10^{30}$ s), significantly restricting the allowed parameter space \cite{Aloisio2025}. Thus, while super-heavy dark matter decay can accommodate a UHE event such as KM3-230213A in specific models, the viable parameter space is strongly restricted by existing multi-messenger observations, and the event provides no positive evidence for a dark-matter origin.

\subsubsection{Primordial black holes and early-universe relics}

UHE neutrinos may also originate from the evaporation of primordial black holes (PBHs), formed in the early Universe. Through Hawking radiation, PBHs can emit energetic particles, potentially producing neutrinos in the PeV--EeV range. Several studies have explored PBH-based interpretations of KM3-230213A, including non-standard scenarios involving modified evaporation, extra dimensions, or connections to dark matter production \cite{Boccia2025,Anchordoqui2025,Singh2025,Aldecoa2026,Jiang2025}. A recent realization involves the decay of super-heavy sterile neutrinos produced during PBH evaporation, yielding an isotropic neutrino flux consistent with KM3-230213A after cosmological redshift \cite{Choi2025}. However, minimal four-dimensional PBH evaporation scenarios are strongly constrained by multi-messenger observations, in particular by the accompanying electromagnetic emission \cite{Airoldi2026}. Thus, PBH interpretations can accommodate KM3-230213A mainly in non-standard or extended scenarios that evade these constraints, but the event provides no positive evidence for a PBH origin.

\subsubsection{Exotic particles and non-standard messengers}

More radical scenarios consider the possibility that the event is not produced by a standard neutrino. In these models, particles from a hidden or dark sector propagate over cosmological distances and interact near the detector, producing signals that mimic neutrino-induced muon tracks \cite{Narita2025,Farzan2025,Dev2025}. Such scenarios can suppress or avoid associated electromagnetic emission and may alleviate the tension between KM3NeT and other experiments.

The local detector environment can affect the sensitivity to some of these scenarios. In particular, the seafloor topography around KM3NeT/ARCA introduces a direction-dependent matter overburden for Earth-skimming particles. An underwater escarpment in the direction of KM3-230213A increases the amount of traversed matter and can enhance the detection probability by up to a factor of $\sim3$ for specific arrival directions \cite{TopographicEnhancement}. This enhancement is a detector-geometry effect and does not, by itself, provide evidence for a non-standard particle interpretation.

Despite their potential to accommodate KM3-230213A while evading some multi-messenger constraints, these scenarios require new particles and interactions, remain highly speculative, and currently receive no positive support from the event itself.

\subsubsection{Electroweak vacuum instability in black-hole mergers}

A more speculative proposal considers UHE-neutrino production during binary black-hole mergers \cite{Sakharov2025}. In this scenario, the strong spacetime curvature reached during the final stages of the merger may induce an electroweak vacuum instability, leading to bubble nucleation and collisions and the subsequent formation of microscopic black holes. Their rapid Hawking evaporation can produce bursts of neutrinos at ultra-high energies. The predicted fluence distribution is strongly non-Gaussian, with rare bright events potentially accommodating KM3-230213A without requiring a correspondingly large diffuse neutrino flux. This may alleviate the tension with the absence of comparable events in IceCube. However, the scenario relies on highly non-standard assumptions about electroweak vacuum dynamics in strongly curved spacetime and remains highly speculative.

Overall, BSM scenarios associated with KM3-230213A span a broad range of physical interpretations, from constraints on modified neutrino propagation to models invoking new particles or unconventional production mechanisms. Their current status is correspondingly diverse: while KM3-230213A provides stringent constraints on some forms of new physics, other scenarios can accommodate the event only under specific and often highly speculative assumptions. At present, none provides compelling evidence for a BSM origin of the event. Nevertheless, KM3-230213A offers a unique opportunity to probe fundamental physics at energies far beyond the reach of terrestrial experiments.

\subsection{Comparison of the proposed scenarios}

The astrophysical and BSM interpretations discussed above span a broad range
of physical assumptions and levels of observational support. While some
astrophysical scenarios can accommodate KM3-230213A within restricted or
model-dependent regions of parameter space, no specific source has been
identified. Other scenarios are strongly constrained by multi-messenger
observations or rely on speculative BSM physics.

The main characteristics, observational constraints, and current status of
the different interpretations are summarized and critically compared in
Table~\ref{tab:scenario_comparison}.

\begin{table}[H]
\caption{Comparison of the main astrophysical and beyond-Standard-Model scenarios proposed in connection with KM3-230213A.}
\label{tab:scenario_comparison}
\centering
\fontsize{5.5}{6.3}\selectfont
\setlength{\tabcolsep}{1.4pt}
\renewcommand{\arraystretch}{1.10}

\begin{tabularx}{\textwidth}{
>{\raggedright\arraybackslash}p{1.35cm}
>{\raggedright\arraybackslash}p{1.45cm}
>{\raggedright\arraybackslash}p{1.10cm}
>{\raggedright\arraybackslash}X
>{\raggedright\arraybackslash}X
>{\raggedright\arraybackslash}X
>{\raggedright\arraybackslash}p{1.25cm}
>{\raggedright\arraybackslash}p{1.25cm}
>{\raggedright\arraybackslash}p{1.45cm}}
\toprule
\textbf{Model or source class} &
\textbf{Neutrino-production mechanism} &
\textbf{Source type} &
\textbf{Main physical requirements} &
\textbf{Main observational constraints} &
\textbf{Tension with IceCube, Auger, or $\gamma$-ray data} &
\textbf{Expected EM signatures} &
\textbf{Role/status for KM3-230213A} &
\textbf{Publication status} \\
\midrule

Galactic origin &
Hadronic interactions of Galactic cosmic rays &
Galactic diffuse / point sources &
Galactic accelerators reaching hundreds of PeV; dense targets &
No compelling Galactic counterpart; diffuse Galactic contribution limited &
Strong $\gamma$-ray constraints &
TeV--PeV $\gamma$ rays &
Disfavored &
Peer-reviewed \cite{GalacticOrigin} \\

\addlinespace
Blazar population &
Photomeson ($p\gamma$) interactions in relativistic jets &
Diffuse population &
Hard proton spectrum ($\alpha_p\leq2$); moderate baryonic loading &
IceCube/Auger UHE limits; Fermi-LAT EGB &
Joint fit compatible; KM3NeT-only in tension with IceCube &
Diffuse $\gamma$ rays &
Compatible &
Peer-reviewed \cite{KM3NeTBlazars} \\

\addlinespace
Individual blazar counterparts &
Hadronic interactions, mainly $p\gamma$, in relativistic jets &
Point / transient sources &
UHE proton acceleration; efficient interactions; high proton power in some models &
Directional uncertainty; no significant association; uncertain source properties &
Transient/$\gamma$-opaque scenarios can alleviate neutrino and $\gamma$-ray constraints &
Radio/X-ray/$\gamma$-ray activity; $\gamma$ rays may be absorbed &
Compatible, but no significant association &
Peer-reviewed \cite{KM3NeTBlazarFollowup,MRC0614Paper,JetRedGiant}; proceedings \cite{PKS0605}; preprint \cite{Kivokurtseva2025} \\

\addlinespace
GRB blastwaves &
Photohadronic interactions of UHE protons &
Diffuse population of transients &
Efficient UHECR acceleration; sufficient baryon loading and ambient density &
Baryon loading--ambient density degeneracy &
Constrained by joint KM3NeT--IceCube--Auger exposure &
Prompt/afterglow emission &
Predicted flux below KM3NeT single-event estimate &
Peer-reviewed \cite{KM3NeTGRB} \\

\addlinespace
Transient sources &
$pp$ and $p\gamma$ interactions of UHECRs &
Transient / diffuse populations &
UHECR acceleration; favorable environment and timescale &
Transient duration/rate; source environment &
Can alleviate IceCube/Auger tension &
Optical/GRB prompt or afterglow emission; may be faint &
Compatible under specific conditions &
Peer-reviewed \cite{Neronov2025,LL-GRBs}; preprint \cite{LFBOTs} \\

\addlinespace
Cosmogenic neutrinos &
UHECR interactions with CMB/EBL &
Diffuse or transient sources &
Proton-rich/light component; hard spectrum/source evolution, or favorable transient/EGMF conditions &
UHECR composition/evolution; EGMF for transients &
IceCube/Auger and $\gamma$-ray limits strongly restrict viable parameter space &
Diffuse or delayed multi-TeV $\gamma$ rays &
Compatible, strongly model-dependent &
Peer-reviewed \cite{CosmogenicKM3NeT,CosmogenicMinimalModels,CosmogenicImplications,CosmogenicPointSource,CosmogenicDelayedGamma}; preprint \cite{CosmogenicMultimess} \\

\addlinespace
Lorentz invariance violation (LIV) &
Modified neutrino propagation &
Independent of source class &
Superluminal propagation from LIV terms &
Propagation of an $E_\nu>100$ PeV neutrino &
Stringently constrained by KM3NeT &
None &
Strongly constrained &
Peer-reviewed \cite{km3net-lvi} \\

\addlinespace
Sterile-neutrino conversion &
Matter-induced sterile-to-active conversion &
Independent of source class &
Sterile flux and new matter interactions &
Different matter path lengths at KM3NeT and IceCube &
Can alleviate KM3NeT--IceCube tension &
None &
Strongly model-dependent; requires non-standard matter effects &
Peer-reviewed \cite{Brdar2026} \\

\addlinespace
Neutrino self-interactions ($\nu$SI) &
Scattering on the C$\nu$B via new light mediators &
Independent of source class &
New scalar/vector mediator; non-zero $\nu$SI coupling &
Spectral distortions/attenuation &
Can mildly alleviate KM3NeT--IceCube tension; strongly constrained by UHE neutrinos &
None &
Strongly constrained / constrained &
Peer-reviewed \cite{He2026}; preprint \cite{Petropavlova2025} \\

\addlinespace
Neutrino--dark matter interactions &
Scattering of UHE neutrinos on dark matter &
Independent of source class &
New mediator and non-zero $\nu$--DM coupling &
Energy-dependent attenuation of the neutrino flux &
Can modify the spectrum and potentially alleviate the KM3NeT--IceCube tension; constrained by UHE-neutrino observations &
None &
Highly model-dependent / constrained &
Peer-reviewed \cite{Bertolez2026}; preprint \cite{Mondol2025} \\

\addlinespace
Super-heavy dark matter decay &
Decay of super-heavy dark matter &
Galactic + extragalactic diffuse &
Super-heavy, ultra-long-lived DM; viable production &
DM lifetime; multi-messenger limits &
Strong neutrino/$\gamma$-ray constraints; viable parameter space remains &
Diffuse $\gamma$ rays; model-dependent GW signatures &
Compatible but strongly constrained &
Peer-reviewed \cite{Borah2025,Kohri2025,Barman2025,Khan2025,Murase205,Aloisio2025}; preprint \cite{Su2025} \\

\addlinespace
Primordial black holes (PBHs) &
Hawking evaporation, directly or via long-lived particles &
Transient / diffuse &
Modified evaporation, extra dimensions, or long-lived secondaries &
PBH abundance, lifetime and multi-messenger limits &
Minimal 4D scenario strongly disfavored; extended models can evade constraints &
$\gamma$ rays for direct evaporation; model-dependent GW signals &
Viable only in non-standard scenarios &
Peer-reviewed \cite{Boccia2025,Anchordoqui2025,Singh2025,Aldecoa2026,Jiang2025,Choi2025,Airoldi2026} \\

\addlinespace
Dark flux / long-lived particles &
Dark-particle propagation, conversion/scattering and decay near detector &
Source-dependent / exotic messenger &
UHE dark flux; weakly interacting/long-lived states &
Interaction strength, decay length and Earth overburden &
Can alleviate KM3NeT--IceCube/$\gamma$-ray constraints via hidden propagation and geometry &
None or strongly suppressed &
Viable in specific BSM models; Viable in specific BSM models; strongly model-dependent &
Peer-reviewed \cite{Narita2025,Farzan2025}; preprints \cite{TopographicEnhancement,Dev2025} \\

\addlinespace
Electroweak vacuum instability in BBH mergers &
Microscopic BH evaporation after vacuum-bubble collisions &
Transient source &
Curvature-induced vacuum instability; microscopic BH formation &
BBH merger rate; UHE-neutrino fluence distribution &
Heavy-tailed fluence can alleviate KM3NeT--IceCube tension &
Possible transient EM counterpart &
Requires non-standard electroweak-vacuum dynamics &
Peer-reviewed \cite{Sakharov2025} \\

\bottomrule
\end{tabularx}
\end{table}

\section{Multi-messenger observations and follow-up studies}

As summarized in Section~3.1, multi-messenger observations provide important constraints on the interpretation of KM3-230213A. Here, we discuss in more detail the dedicated electromagnetic follow-up campaigns performed within the event localization region. These observations span the gamma-ray, X-ray, optical, and radio bands and have not identified an unambiguous counterpart. In the following, we focus on the main results from gamma-ray studies and radio follow-up observations.

\subsection{Gamma-ray and high-energy constraints}

The production of UHE neutrinos through hadronic interactions is generally expected to be accompanied by high-energy gamma rays. These photons, produced via pion decay, do not propagate freely over cosmological distances but interact with the extragalactic background light (EBL), initiating electromagnetic cascades that redistribute the energy to the GeV--TeV range. Recent studies have investigated the expected gamma-ray counterpart of KM3-230213A \cite{Fang2025}. In this framework, UHE photons ($E_\gamma \sim 400$ PeV) initiate electromagnetic cascades via pair production and inverse Compton scattering, producing a spectrum largely independent of the initial emission and peaking in the $\sim 100$ GeV to multi-TeV range, depending on the source redshift. The observability of this emission depends on the intergalactic magnetic field: for weak fields ($B \lesssim 10^{-14}$ G), the signal may appear point-like on timescales of years and be detectable by instruments such as Fermi-LAT and ground-based gamma-ray observatories, whereas stronger fields can delay and suppress the emission, spreading it over larger angular scales and longer timescales.

Very-high-energy gamma-ray follow-up observations of the KM3-230213A region have been performed with VERITAS \cite{VERITAS}. No significant excess was detected within the 90\% localization region, and upper limits at the level of $\sim1.8\%$ of the Crab Nebula flux above 550 GeV were derived. This non-detection is consistent with several scenarios, including intrinsic gamma-ray absorption within the source, attenuation by the EBL, or transient emission occurring outside the observation window. These results place important constraints on hadronic models linking the neutrino to TeV-bright sources. A complementary study based on long-term Fermi-LAT observations has also searched for the expected electromagnetic cascade signal associated with the neutrino event, finding no significant gamma-ray counterpart in the GeV–TeV range \cite{Crnogorevic2025}. This non-detection places additional constraints on hadronic scenarios and on the source environment, reinforcing the challenges in identifying a clear multi-messenger counterpart.

\subsection{Radio follow-up and candidate counterparts}

\begin{figure}[htbp]
\centering
\includegraphics[width=0.55\textwidth]{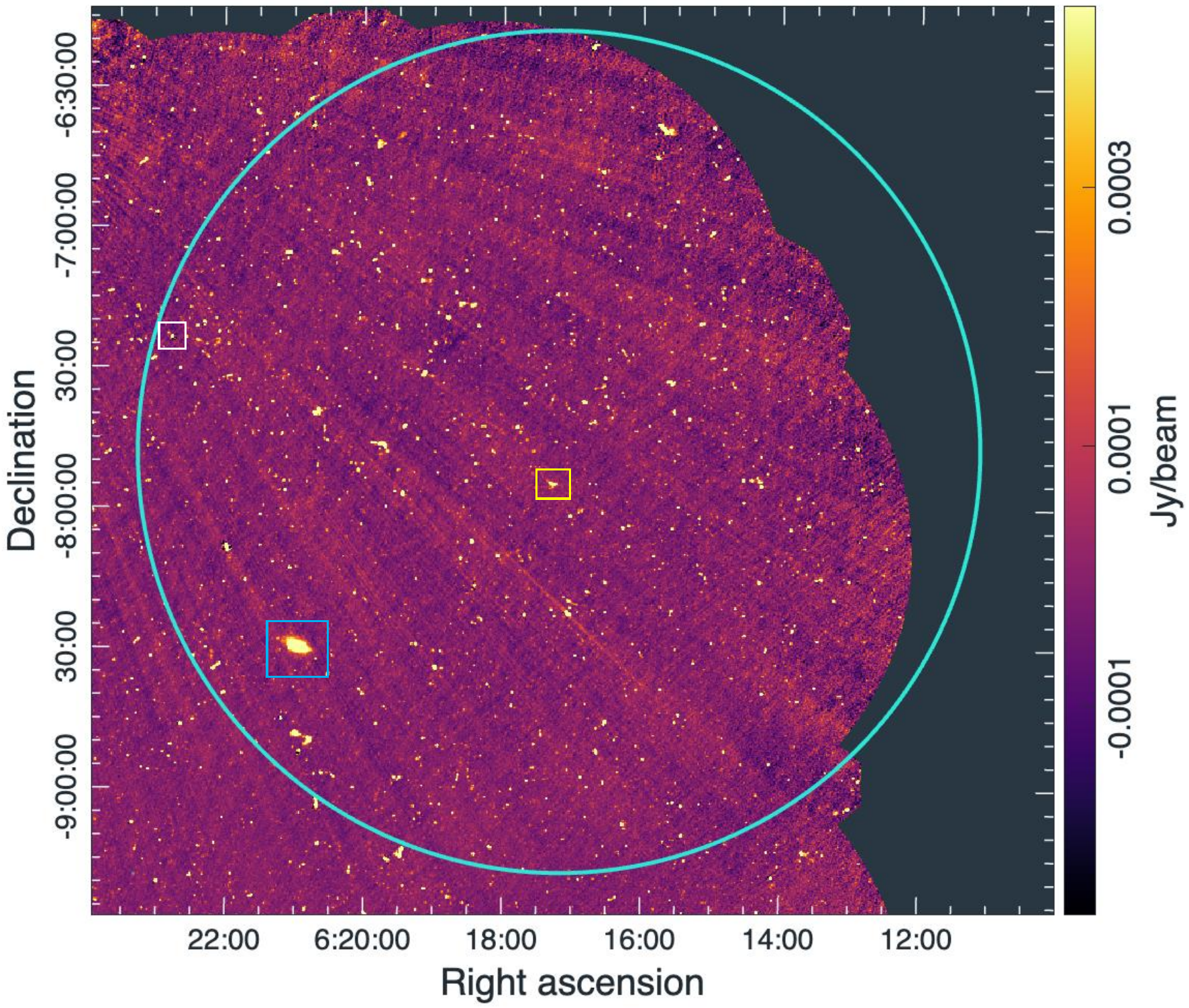}
\caption{Radio map of the KM3-230213A localization region obtained with ASKAP. The field contains a large number of radio sources, including several candidate AGNs, illustrating the difficulty of identifying a unique counterpart. Figure adapted from \cite{ASKAPradio}.}
\label{fig:askap_map}
\end{figure}

Radio observations provide a complementary probe for identifying potential counterparts within the localization region. Dedicated follow-up studies using ASKAP together with VLASS observations have revealed a large population of radio sources within the error region \cite{ASKAPradio}. As illustrated in Fig.~\ref{fig:askap_map}, the radio field is characterized by a high source density. More than a thousand radio sources are detected within the 68\% containment region, among which a subset exhibits properties consistent with active galactic nuclei, including flat spectra and variability. Seventeen candidates, such as compact variable sources and galaxies hosting jet activity, emerge as potentially interesting counterparts. However, no single source can be unambiguously associated with the neutrino event. The large neutrino localization region, combined with the high density of radio sources, makes it difficult to identify a unique counterpart. This illustrates the challenges of counterpart identification for UHE neutrinos and suggests that multiple source classes may contribute to the observed signal.

These results emphasize the importance of coordinated multi-messenger observations in constraining the origin of UHE neutrinos. Future experimental efforts will be crucial to further clarify the nature of events such as KM3-230213A.

\section{Future experimental prospects}

The interpretation of KM3-230213A highlights the need for next-generation experiments capable of probing UHE neutrinos with larger effective volumes and complementary detection techniques. Space- and balloon-based experiments such as PBR, a pathfinder for POEMMA, are designed to perform target-of-opportunity observations of transient astrophysical events and to search for neutrinos in the energy range $10^8 \lesssim E_\nu / \mathrm{GeV} \lesssim 10^{10}$ \cite{Olinto2025}. These instruments are particularly well-suited to test scenarios involving transient sources or BSM messengers that could accommodate the KM3NeT detection without corresponding signals in IceCube or the Pierre Auger Observatory. Sensitivity estimates indicate that events similar to KM3-230213A may be within reach under favorable conditions.

Complementary approaches are pursued by large-scale radio detector arrays such as the Askaryan Radio Array, which exploit coherent radio emission from neutrino-induced showers in ice to achieve the large detection volumes required at EeV energies \cite{Muzio2025}. Recent analyses show that sensitivities above $\sim 3$ EeV are approaching the level needed to test the flux implied by the KM3NeT event, while also paving the way for future instruments such as the radio component of IceCube-Gen2 \cite{IceCubeGen2} and RNO-G \cite{RNOG}. Other complementary approaches include GRAND, a large-scale radio array designed to detect air showers induced by Earth-skimming UHE tau neutrinos \cite{GRAND}, as well as TAMBO and Trinity, which target Earth-skimming tau neutrinos using, respectively, a deep-valley detector array and an imaging atmospheric Cherenkov technique \cite{TAMBO,Trinity}. Taken together, these developments suggest that the next generation of detectors will play a crucial role in determining whether KM3-230213A represents the extreme high-energy tail of the known astrophysical neutrino flux or the first indication of a new UHE neutrino component.

In this context, the observation of near-horizontal UHE events such as KM3-230213A has also stimulated the development of new detector concepts specifically optimized for this geometry. Recent proposals explore sparse and planar configurations designed to enhance sensitivity to horizontal muon tracks, highlighting a complementary strategy to traditional large-volume neutrino telescopes \cite{POLARIS}.

\section{Conclusions}

The detection of KM3-230213A marks a significant milestone in UHE neutrino astronomy. Although based on a single event, it demonstrates the capability of KM3NeT to probe the neutrino sky beyond 100~PeV and opens the way toward population studies at the highest energies.

Despite extensive multi-messenger follow-up efforts, no unambiguous counterpart has been identified, and the origin of the event remains uncertain. Conventional astrophysical sources such as AGNs and GRBs remain viable candidates, although they require either extreme source conditions or electromagnetically faint emission, while the cosmogenic neutrino hypothesis provides a natural explanation within standard physics under specific assumptions on the composition and evolution of UHECR sources. More speculative interpretations involving physics beyond the Standard Model, including modified neutrino propagation, dark matter decay, or exotic particle scenarios, have also been explored, but are not required by current data and remain strongly model dependent.

The KM3-230213A event has stimulated extensive activity in the multi-messenger astrophysics community, highlighting both the opportunities and challenges of UHE neutrino observations. The lack of a clear electromagnetic counterpart and the tension with IceCube and Auger constraints remain key issues, as emphasized in recent multi-messenger studies \cite{multi-messengerReview}. Future increases in detector exposure, together with next-generation experiments and coordinated observations, will be crucial to determine whether events of this kind represent the high-energy tail of known astrophysical processes or the first indication of new physical phenomena.

\section*{Acknowledgments}

The author would like to thank the KM3NeT Publication Committee for valuable comments and feedback that helped improve the manuscript.

\section*{Conflicts of Interest}

The author declares no conflict of interest.

\small

%=====================================
% References
%=====================================

\end{document}